\documentclass{appolb}
\usepackage{graphicx}
\usepackage{amsmath} 
\usepackage{amssymb} 
\usepackage{slashed} 
\usepackage{bigstrut} 
\usepackage{mathtools} 
\usepackage{multirow}
\usepackage{color}
\usepackage{hyperref}
\usepackage{url}
\usepackage[square,numbers,sort&compress]{natbib}
\begin{document}
\title{
  BSM and SM signals and backgrounds \\
  in Far-Forward Experiments at the LHC 
\thanks{Presented at ``Diffraction and Low-$x$ 2022'', Corigliano Calabro (Italy), September 24-30, 2022.}%
}
\author{M.~V. Garzelli
  \address{Universit\"at Hamburg, II Institut f\"ur Theoretische Physik,\\
    Luruper Chaussee 149, D-22761 Hamburg, Germany}  
\\[3mm]
}
\maketitle
\begin{abstract}
Two far-forward experimental systems are currently taking data du\-ring Run 3 at the Large Hadron Collider (LHC): FASER~+~FASER$\nu$
and 
SND@LHC.
They are sensitive to some classes of beyond-the-Standard Model (BSM) particles,
muons and neutrinos produced in the ATLAS interaction point (IP) and propagating for several hundred meters along the tangent to the accelerator beamline, up to
the caverns where they are respectively located, in opposite directions with respect to the IP.
Proposals are being prepared to extend these experiments to bigger ones during the HL-LHC phase. Building a Forward Physics Facility (FPF) capable of hosting a number of far-forward experiments characterized by different detection techniques, kinematical acceptance and purpose, is a possibility also under discussion. In this contribution I discuss some of the BSM and SM signals and backgrounds at the FPF,
mainly focusing on QCD-related aspects.
\end{abstract}
  
\section{Introduction}
The fact that hadronic collisions at colliders produce an abundant number of neutrinos, especially in the forward region, is known since long~\cite{DeRujula:1992sn, Park:2011gh}. However, the detection of these neutrinos and a systematic investigation of
the associated Physics reach
has not been the subject of specific research efforts and investments, until the proposal in very recent years of two expe\-riments, SND@LHC~\cite{SNDLHC:2022ihg} and FASER$\nu$~\cite{FASER:2019dxq}. Their main purpose is detecting the most forward of the neutrinos produced in one of the IPs
at the 
LHC
and propagating along the tangent to the accelerator arc, thanks to their interactions with the nuclei of sui\-ta\-ble detectors. These experiments were built 
only recently and data taking has just begun 
du\-ring Run~3. They are located in two service caverns along the tangent to the LHC proton beamline, at $\sim$~480~m from the ATLAS interaction point, in opposite directions. FASER$\nu$ occupies the UJ12 cavern, whereas SND@LHC
the UJ18 one. In addition to neutrinos, they are sensitive to muons, mainly considered as a background, to dark matter (DM) elastically or inelastically scattering in the detector and to long lived particles (LLPs)/feebly interacting particles (FIPs) decaying either within the Standard Model (SM) or beyond it (BSM). A second experiment, devoted to LLP/FIP and DM searches, called FASER~\cite{FASER:2022hcn}, is also located in UJ12 and positioned downstream of FASER$\nu$. In fact, not only the flux of neutrinos but even the flux of light LLPs/FIPs and light DM particles, which can be produced at the IP via a number of mechanisms, 
is expected to be large in the forward direction~\cite{FASER:2018eoc}.

Additionally, in view of Run 4, upgrades of all these experiments are under study, accompanied by the investigation of the possibility of building a dedicated Forward Physics Facility (FPF)~\cite{MammenAbraham:2020hex, Anchordoqui:2021ghd}, large enough to probably host at least some of the upgraded detectors and additional experiments. One of the limitations of the present setups is the smallness of the caverns in use, which does not allow to host large-mass detectors. The FPF, allowing for bigger detectors, would be able to dramatically increase the statistics accumulated during Run 3. However, its perspective location being further from the ATLAS IP, i.e. at $\sim$~600~m, compared to the baseline $\sim$~480~m of the present experiments, implies that the minimal neutrino pseudorapidities that can be explored with a decent statistic are still quite large and similar to those explored during Run 3, i.e. $\eta_\nu \gtrsim 7$. The lack of co\-ve\-rage of lower $|\eta_\nu|$ could be partially resolved by complementing the  FPF detectors with other ones in different $\eta_\nu$ ranges at different sites (see e.g. the proposal of an Advanced SND-Near detector, to be positioned at pseudorapidities $4~<~\eta_\nu~<~5$, i.e. certainly outside the FPF, complementing the measurements of the Advanced SND-Far detector, FPF successor of SND@LHC, 
at pseudorapidities $7.2 < \eta_\nu < 8.4$) and/or by making some of the FPF detectors working in timing coincidence with the ATLAS ex\-pe\-ri\-ment, as sketched in ref.~\cite{Anchordoqui:2021ghd}. Single neutrinos are not di\-stin\-guishable and just contribute to missing energy in ATLAS, as well as in the other main LHC detectors. However the ATLAS apparatus would be sensitive to a number of other particles produced at central pseudorapidities in the same event leading to at least a neutrino detected by FPF detectors. 

In the following we review some of the most important signals and backgrounds for the aforementioned far-forward experiments, of interest for the BSM and/or the SM physics programs. In particular, we focus on QCD-related aspects. Complementary and/or additional information can be found in the recent Snowmass FPF reports published in ref.~\cite{Anchordoqui:2021ghd} and~\cite{Feng:2022inv} and further references therein. 

\section{BSM physics}

\begin{figure}[h]
\begin{center}
\includegraphics[width=0.50\textwidth]{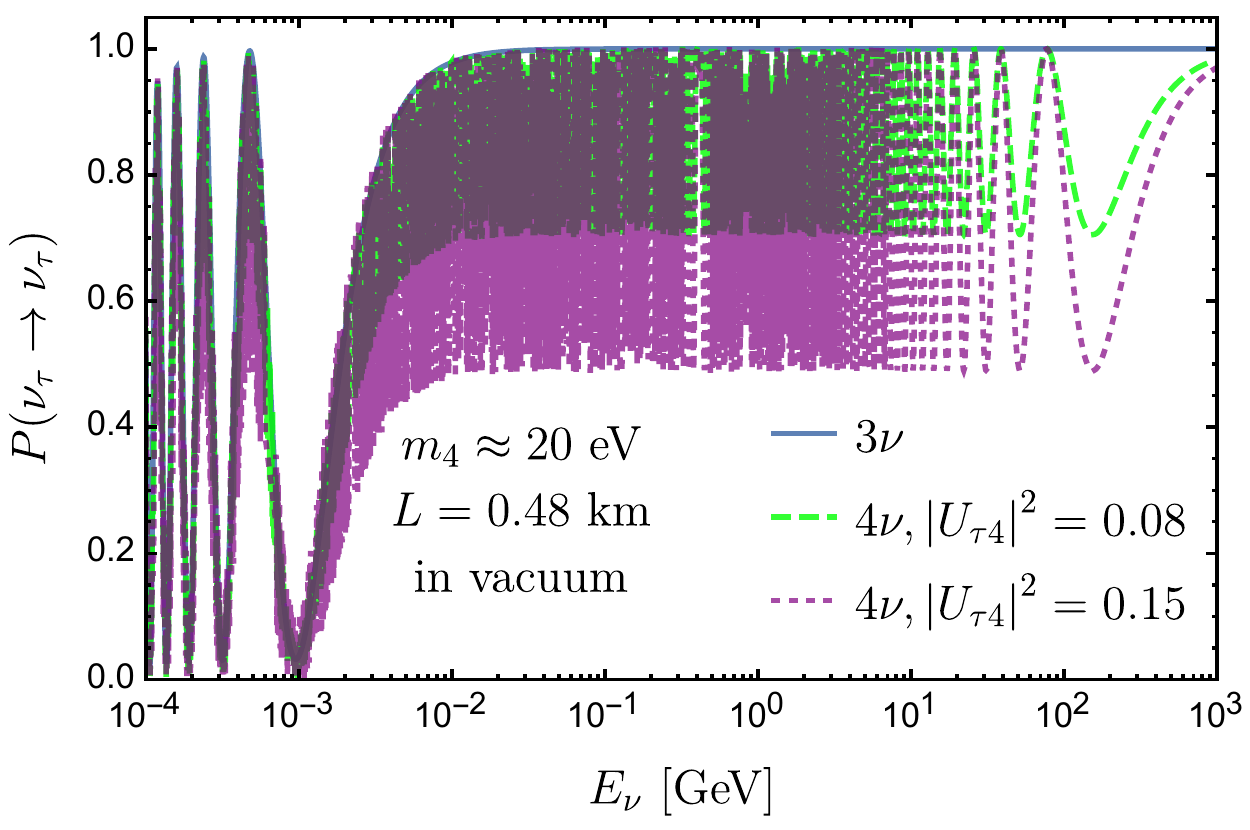}
\includegraphics[width=0.48\textwidth]{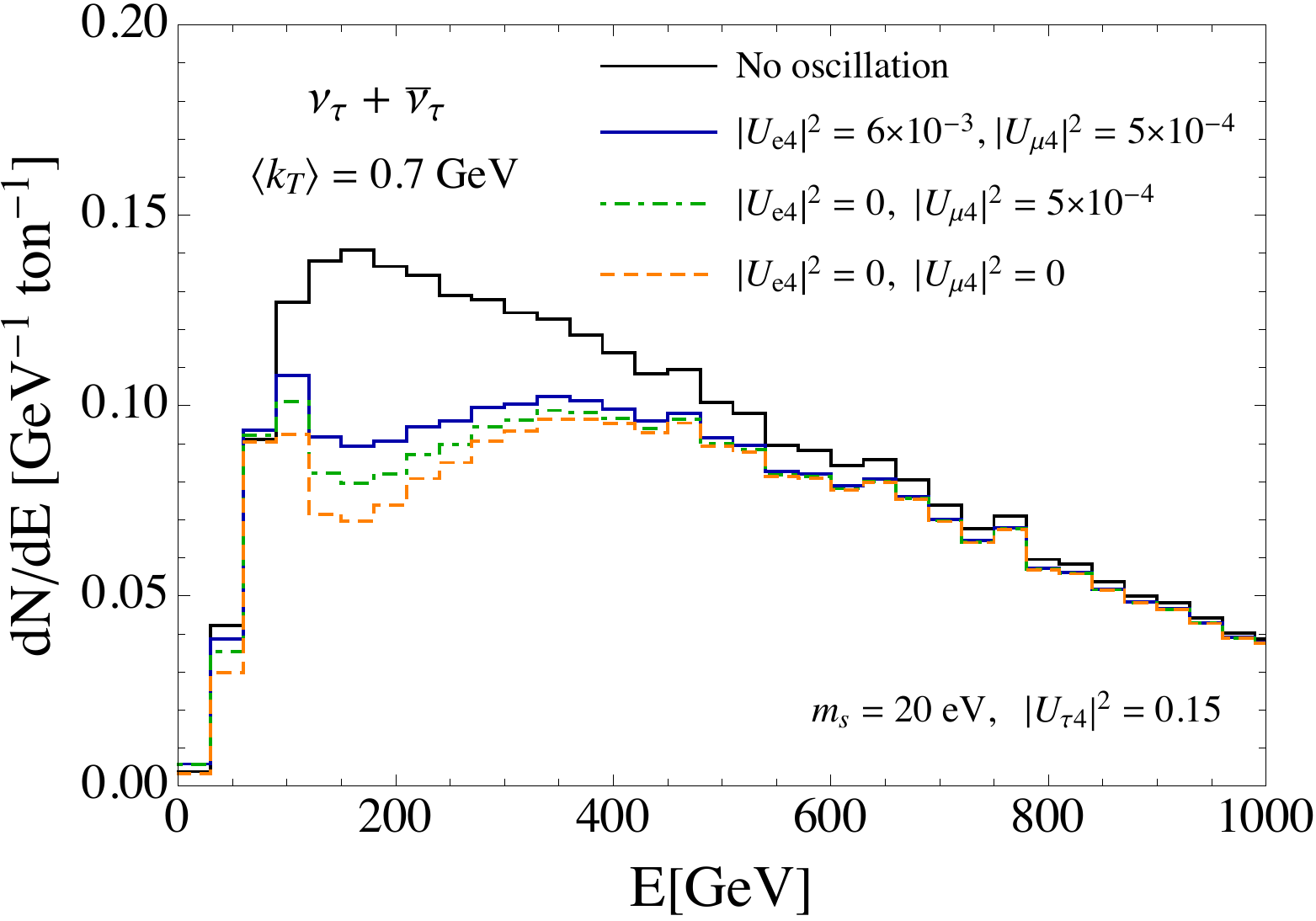}\\
\caption{\label{fig:oscilla}
  $\nu_\tau$ survival probability as a function of $E_\nu$
at a baseline of 480~m for a 3-active and a 3-active + 1-sterile neutrino scenario with mass eigenstate $m_4$ value and mixing parameter $U_{\tau 4}$ values as specified in the plot ({{\it left}}).   
Oscillation of $\nu_\tau$ in heavy sterile neutrinos ({$m_4 \sim$ 20 
eV})  can be probed 
by looking at deficit or excess in the observed ($\nu_\tau$ + $\bar{\nu}_\tau$)
event spectrum ({\it{right}}), once the QCD uncertainties on this spectrum (not shown in the plot) are well under control.
See ref.~\cite{Bai:2020ukz}, from which these plots are extracted, for more detail.
}
\end{center}
\end{figure}

A first BSM investigation that can be done with all present and future
far-forward neutrino experiments at the LHC is the study of active to sterile neutrino oscillations. Neutrinos are massless in the minimal SM. However, a number of experiments have unvealed neutrino oscillation phenomena, which point to the fact that at least three neutrino mass eigenstates e\-xist and raise the question on the origin of their masses and how the SM has to be extended to accomodate them~\cite{Giunti:2022aea}. The baseline of several hundred meters of the far-forward neutrino experiments at the LHC is not appropriate for disentangling active to active neutrino flavour oscillations, which are averaged out at these characteristic distances,
but can be adopted for studying oscillations of active neutrinos, in particular $\nu_\tau$, to sterile neutrinos with masses of the order of 10's of eV~\cite{Bai:2020ukz},
as shown in Fig.~\ref{fig:oscilla}.

A second class of BSM investigations that can be done with the far-forward detectors is the search for LLPs/FIPs and dark matter (DM). Due to their cha\-rac\-te\-ri\-stic lifetimes, various kinds of LLPs coupled to SM particles and produced at the IP  
could escape the traditional large massive LHC detectors, propagate and decay in the far-forward
detectors, or in their path towards the latter. These decays could give rise to either SM particles, 
or to DM particles, which could further scatter with the material of these detectors, giving rise to interesting signals, sometimes mimicking SM ones. Various kinds of mediators to Dark Sectors (DS) have been con\-si\-de\-red: dark bosons, dark scalars, heavy neutral leptons, and related benchmark models have been proposed by the CERN Physics Beyond Colliders study group.
The sensitivity of present and future far-forward experiments in case of these scenarios has been assessed in a number of works (see ref.~\cite{Anchordoqui:2021ghd, Feng:2022inv} and refs. therein). The most studied case is the dark photon $A^\prime$, either decay\-ing into DM, or only kinetically mixed with the SM photon $\gamma$, which leads to $\gamma \rightarrow A^\prime \rightarrow \gamma$ transitions.
Millicharged (MC) particles, cor\-re\-spon\-ding to the particular case $m_A^\prime = 0$, have also been the subject of dedicated efforts.
The possibility of including a specific detector (FORMOSA~\cite{Foroughi-Abari:2020qar}) for MC particles at the FPF has been investigated, as well as the one of detecting MC particles by means of multipurpose detectors, such as FLArE, also useful for investigating neutrino interactions and DM scattering. A\-no\-ther class of LLPs, the axion-like particles, could just be coupled to the SM $\gamma$'s,~fermions or gluons, without need of introducing any DS. On the other hand, light DM particles could be produced already at the IP and propagate to the detectors in BSM scenarios which do not need any LLP/FIP mediator.

Controlling the SM backgrounds is 
a key requirement for a successful BSM program.  This involves, among other objectives, an accurate charac\-te\-ri\-za\-tion of the muon fluxes reaching the detectors, as well as of the neutrino fluxes.
Additionally, with the purpose of unvealing DM interactions with the nuclear media of the detector, one needs a good control of the $\nu e$, as well as the $\nu N$ ($\nu A$) backgrounds, including the role of nuclear effects due to the detector composition. 
Focusing on particularly low recoil energies ($E_{rec} \lesssim$~1~GeV) for recoil angles $\theta_{rec} \sim 50$~mrad with specific kinematics cuts, might help to drastically reduce these two sources of background~\cite{Batell:2021blf}.

\section{SM physics}

A wide SM physics program can be conducted at far-forward neutrino~experiments, in the hypothesis that BSM effects like those mentioned in the~previous section are either absent or limited and well understood. Such a program is focused on two main topics, determining neutrino fluxes and~cross sections, and related aspects. 
The detectors are sensitive to the convolution~of these two elements, and not to each of them separately. The~strate\-gy of building ratios of selected observables can be employed to cancel the effects of either the fluxes or the cross sections, allowing to access the informa\-tion of interest at each time. The neutrino beams impinging on the detectors have collider-frame energies up to several TeV. Further neutrino beams with such~high energies are not available on the Earth nowadays. The far-forward neutrino experiments at the LHC, in this respect, offer a unique opportunity.

Above the GeV scale, 
neutrino inelastic interactions
with nucleons occur via deep inelastic scattering (DIS), the physics of which is known~up~to approximate N$^3$LO accuracy in pQCD
while efforts towards N$^4$LO have started~\cite{Basdew-Sharma:2022vya}. The presence of nuclei, instead of just nucleons, brings~in additional corrections and increases the uncertainty of the predictions. In any case, we can reasonably argue that, nowadays, our knowledge of neutri\-no DIS cross sections is more advanced than the one of far-forward neutri\-no fluxes. Non-perturbative QCD effects, not well understood in the present QCD theoretical framework, play a crucial role in determining these fluxes. In present days, non-perturbative QCD effects are not derived by first principles, but described by phenomenological models, involving a number of parameters fixed by fits of experimental data. The uncertainties associated both with the models themselves, and with the fits of their parameters, are large and the criteria to assign/evaluate these uncertainties are still subject of intense debate.
As a consequence, pre\-dic\-ting the values of far-forward neutrino fluxes, as well as their uncertainties, represents one of the key theory challenges at far-forward experiments. The fluxes of neutrinos from the decay of light mesons produced at the IP are truly dominated by non-perturbative QCD. On the other hand, for the fluxes of neutrinos from~the decay of heavy-flavoured mesons, a pQCD treatment can be applied, consi\-dering that the charm and bottom mass values are well above the $\Lambda_{\mathrm{QCD}}$ value. However, the relative smallness of these masses, especially the charm mass, larger than $\Lambda_{\mathrm{QCD}}$ but not too large, makes non-perturbative effects also quite relevant. 
If we can disentangle fluxes from cross-sections at the FPF, we will be able to learn about forward light- and heavy-meson production. This is useful for a number of applications, not limited to QCD, but ranging from high-energy
cosmic-ray
physics to neutrino astronomy.

\begin{figure}
  \begin{center}
\includegraphics[width=0.49\textwidth]{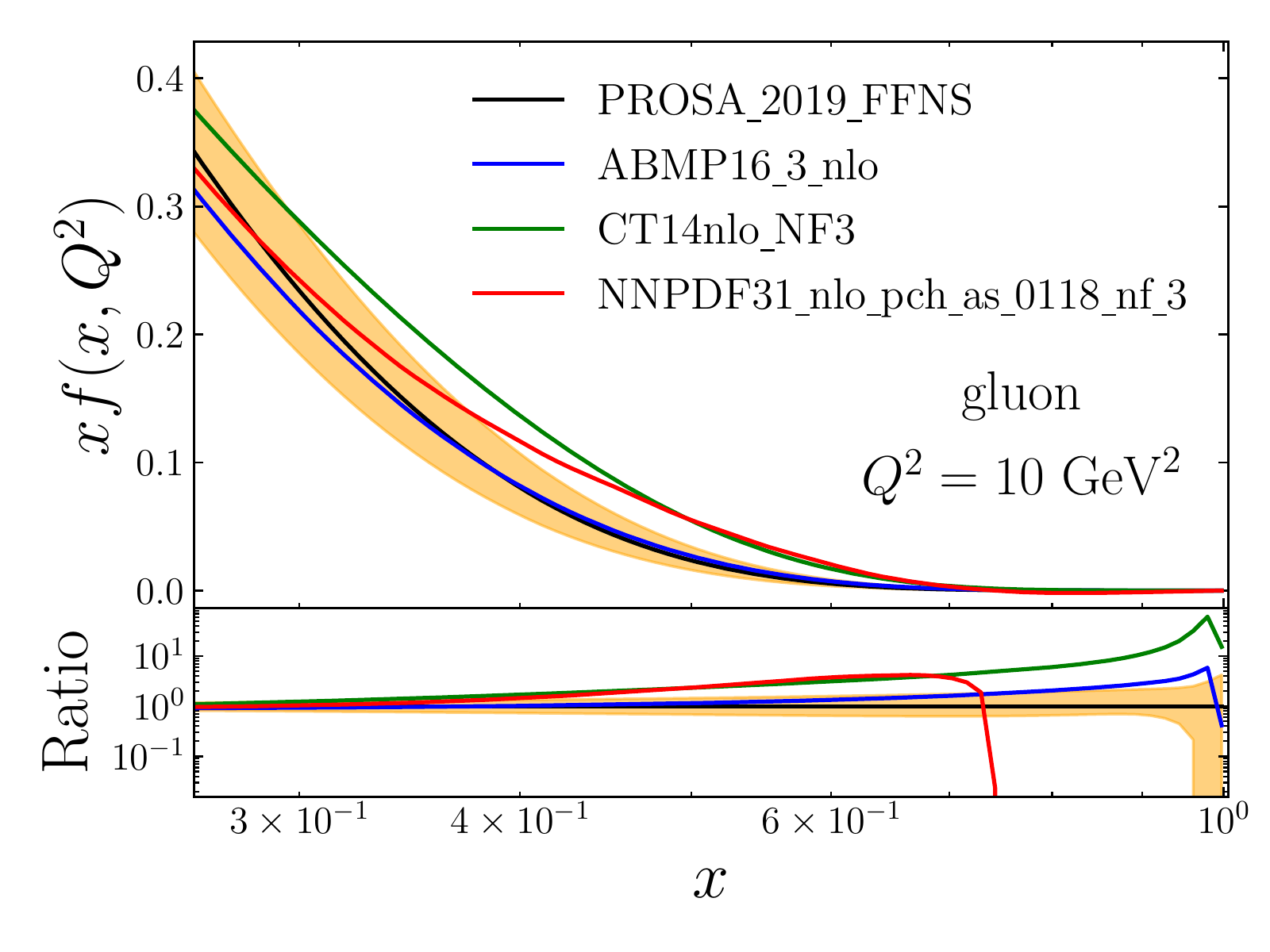}      
\includegraphics[width=0.40\textwidth]{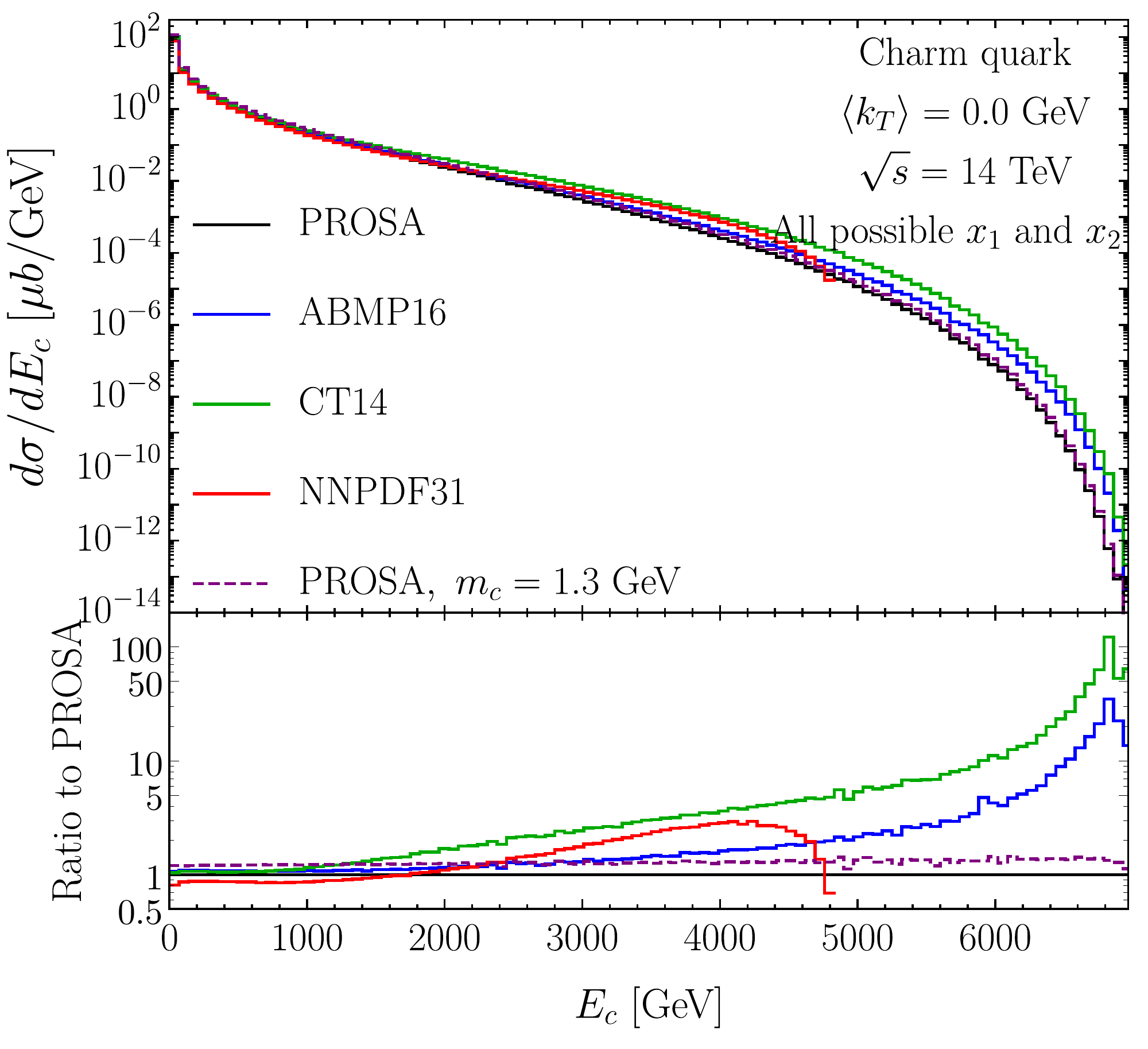}
  \caption{\label{fig:largex}
    Gluon PDFs $xf(x,Q^2)$ for $Q^2=10$ GeV$^2$ 
    for the PROSA 2019~\cite{Zenaiev:2019ktw}
    fit (black curve with orange uncertainty band). The  ABMP16~\cite{Alekhin:2018pai}, CT14~\cite{Dulat:2015mca} and NNPDF3.1~\cite{NNPDF:2017mvq} central gluon NLO PDFs with 3 active flavours are also shown ({\it{left}}). 
    Charm energy distributions at NLO from $pp \rightarrow c\bar{c} + X$
    using as input these same PDFs ({\it{right}}). 
    The lower insets show ratios relative to the central PROSA PDF.
    See ref.~\cite{Bai:2021ira}, from which these plots are extracted, for more detail.
}
\end{center}
\end{figure}

\begin{figure}
\begin{center}
\includegraphics[width=0.48\textwidth]{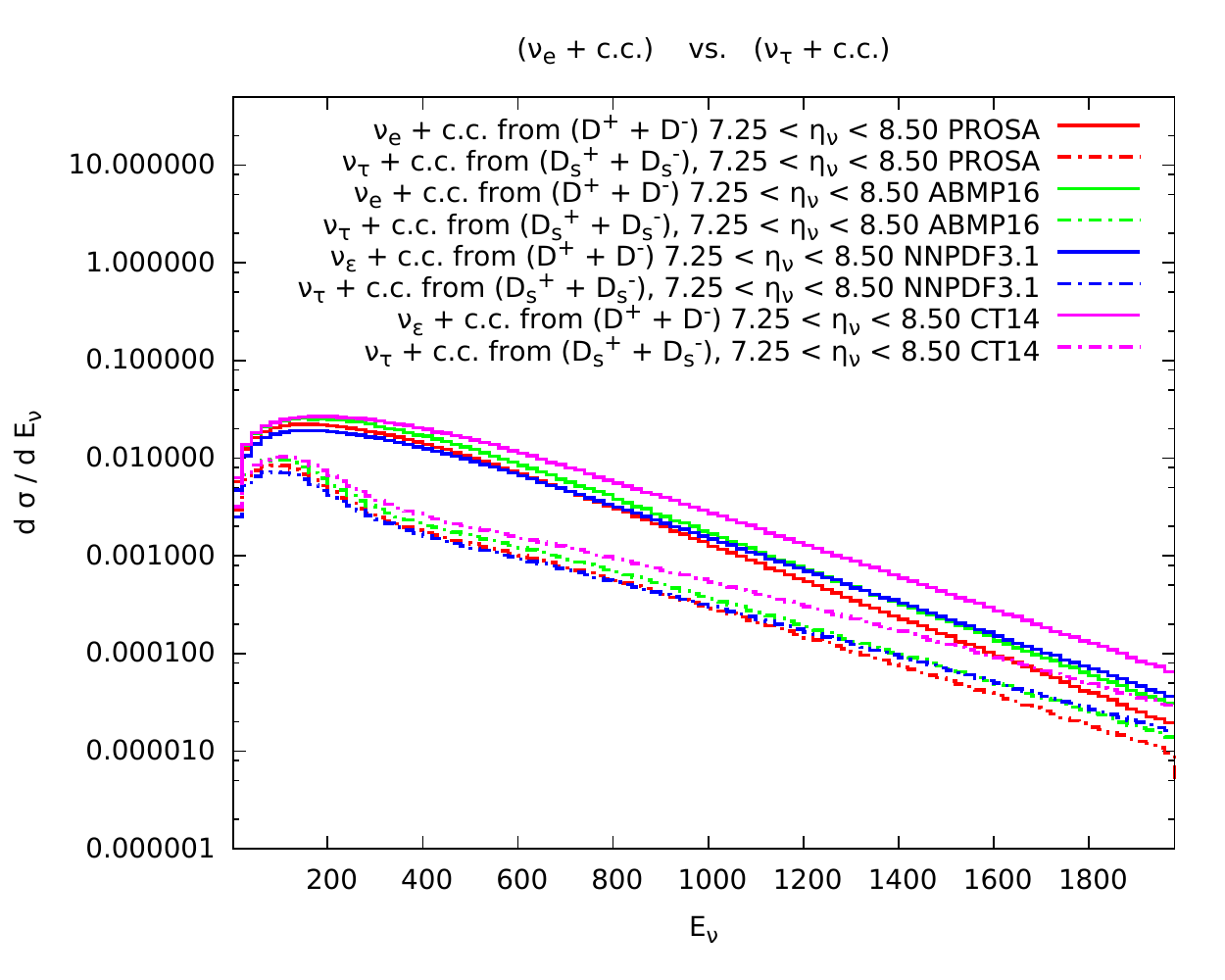}
\includegraphics[width=0.48\textwidth]{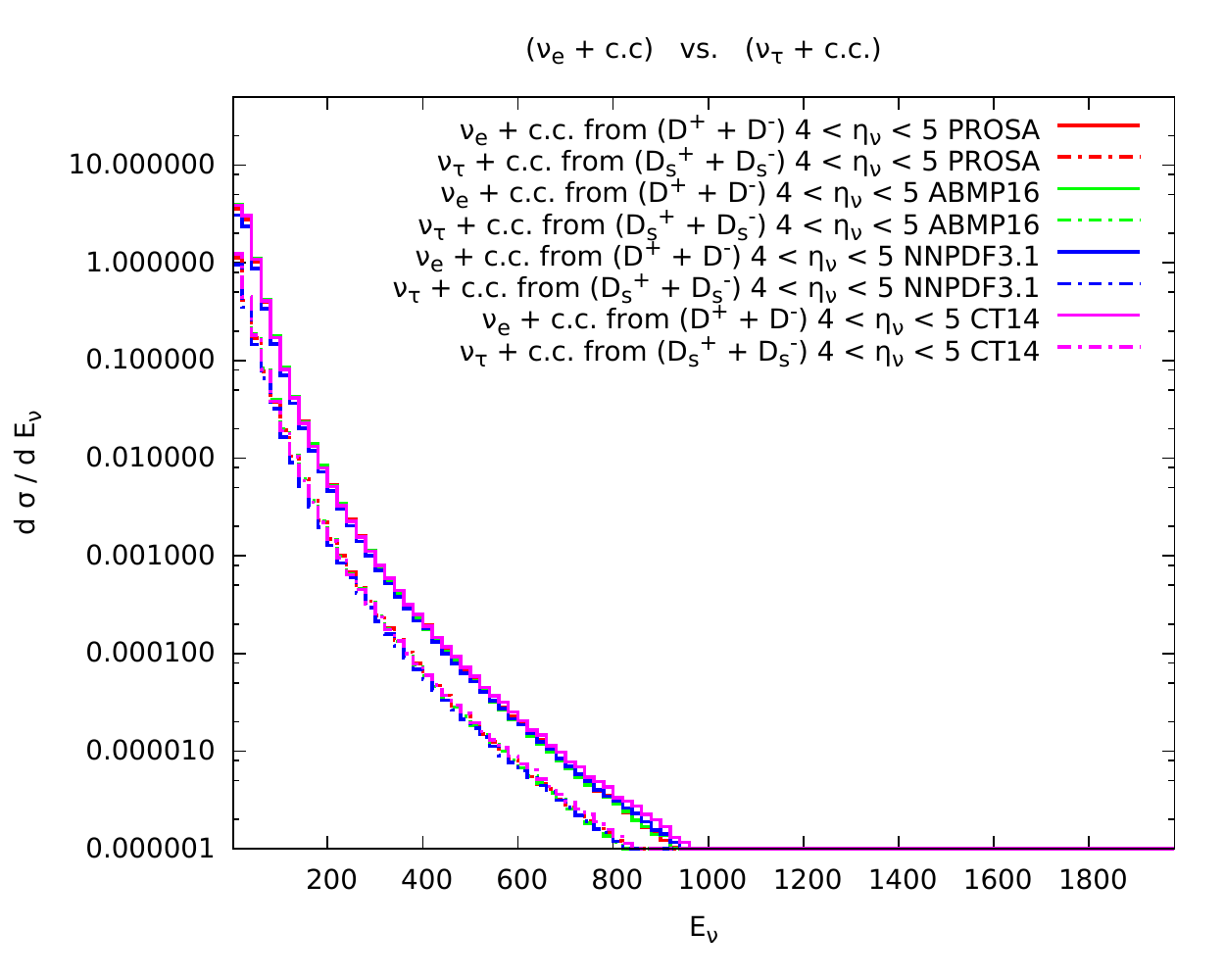}
\caption{
  \label{fig:nuenutau}
 Predictions for $(\nu_e +\bar{\nu}_e)$ fluxes from $D^\pm$ decays and $(\nu_\tau + \bar{\nu}_\tau)$ fluxes from $D_s^\pm$ decays in the
$7.2 < \eta_\nu < 8.4$ (left) and $4 < \eta_\nu < 5$ (right)
  pseudorapidity ranges, using the different central PDF sets already considered in Fig.~\ref{fig:largex}. See ref.~\cite{Bai:2022jcs} for more detail. 
}
\end{center}
\end{figure}

\begin{figure}
\begin{center}
\includegraphics[width=0.43\textwidth]{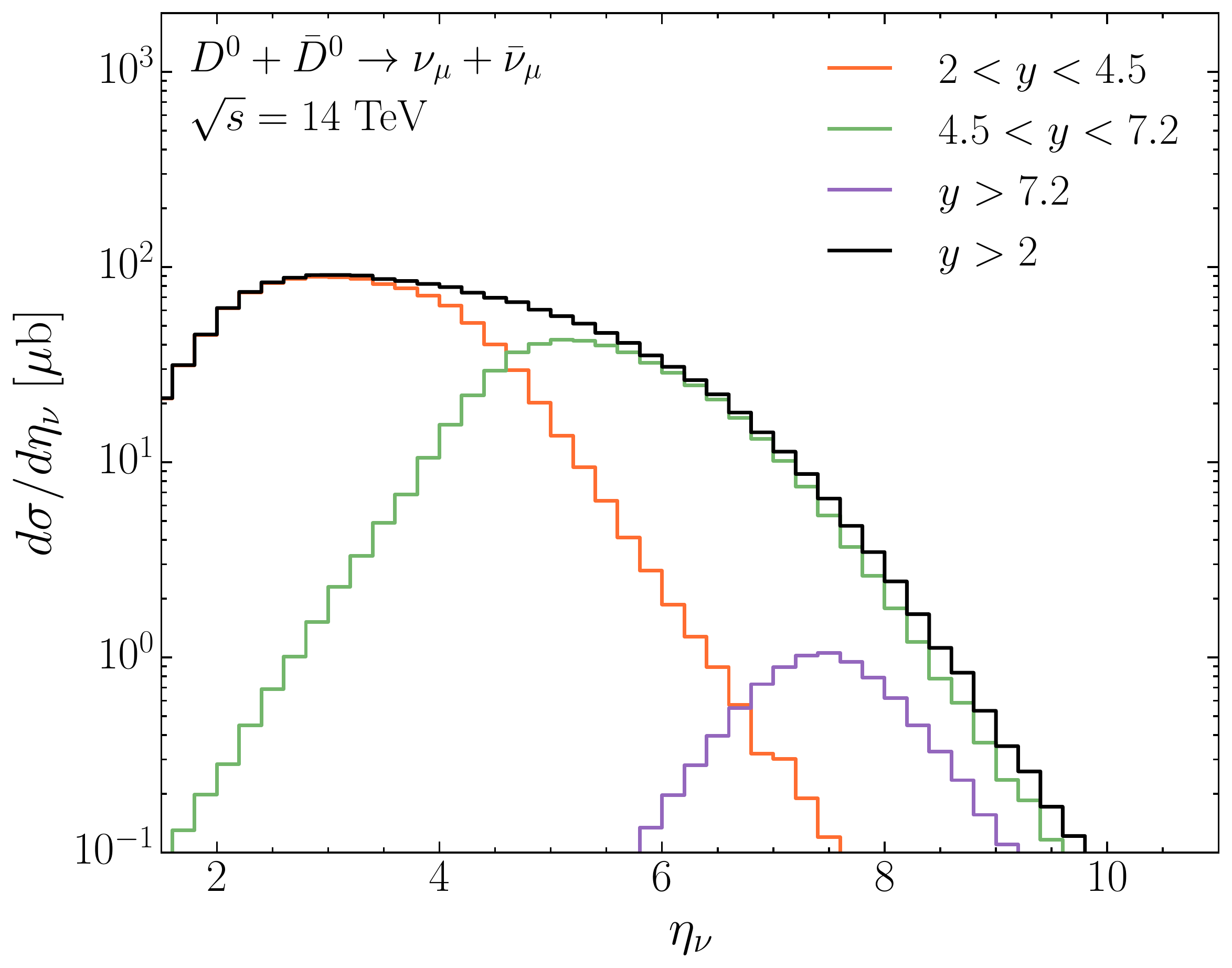}
\includegraphics[width=0.44\textwidth]{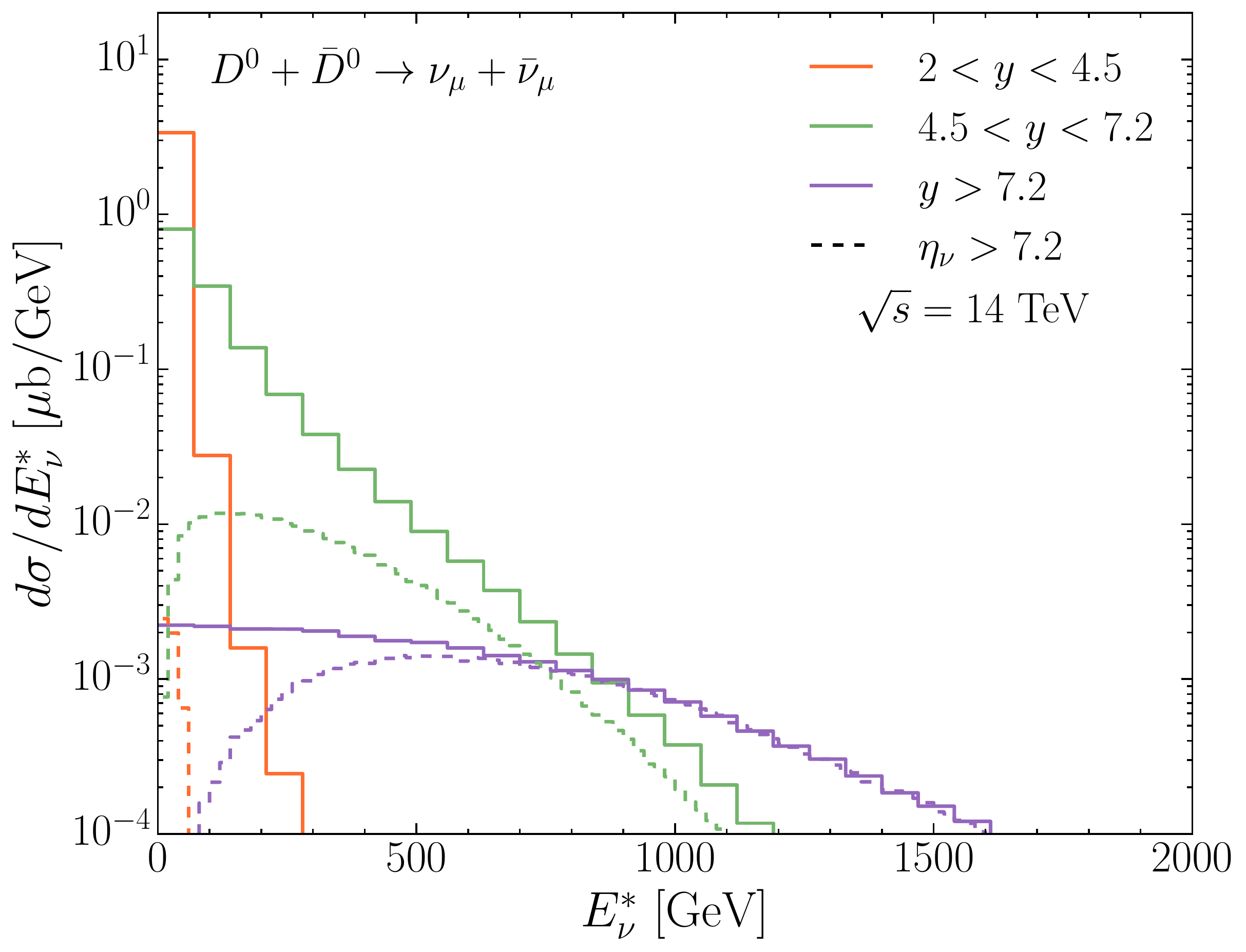}
\caption{Predictions for the pseudorapidity ({\it {left}}) and energy ({\it {right}}) spectrum of ($\nu_\mu$~+~$\bar{\nu}_\mu$)
  from $pp \rightarrow (D^0 + \bar{D}^0) + X$ with
  charmed meson rapidity $y$ in different intervals. 
The dashed histograms show the energy spectra from the corresponding charmed meson $y$ intervals that give neutrinos and antineutrinos with $\eta_\nu > 7.2$.  
See ref.~\cite{Bai:2022xad}, from which these plots are extracted, 
for more detail. 
\label{fig:promptnu}
}
\end{center}
\end{figure}

One of the most appealing QCD outcome of this program would be better constraining the gluon and sea quark PDFs in the low and large longitudinal momentum fraction $x$ domains, to which forward heavy-flavour production is sensitive~\cite{Zenaiev:2019ktw, Begel:2022kwp}. In the left panel of Fig.~\ref{fig:largex},
the large-$x$ gluon distribution is shown as a function of $x$ for different NLO PDF sets. Especially for $x > 0.4$, the best-fit results of different PDF sets turn out to be quite different among each other and they are not always covered by the uncertainty associated to each single set. The variation pattern of the considered large-$x$ gluon PDFs is directly reflected in the behaviour of the corrresponding energy distributions of charm quarks produced at the IP, that, at energies $E_c >  2$ TeV, show differences among each other similar to those observed for the PDFs, as follows from comparing the right- and left-hand side of Fig.~\ref{fig:largex}. In turns charm quarks fragment into $D$-hadrons, which decay, producing a neutrino flux. We can then conclude that the energy distribution of charm-induced neutrinos above~1~TeV is very sensitive to large-$x$ PDFs. This is also visible in Fig.~\ref{fig:nuenutau}, where, for the same PDF sets as above, the energy distributions of $\nu_\tau + \bar{\nu}_\tau$ from $D_s^\pm$ decay and of $\nu_e + \bar{\nu}_e$ from $D^\pm$ decay are shown in two different neutrino pseudorapidity ranges, $7.2 < \eta_\mu < 8.4$ and $4 < \eta_\nu < 5$, corresponding to the coverages of the Advanced SND-Far and Near detectors, foreseen for the Run 4 upgrade of SND@LHC. From the plots it is evident that, the larger are the neutrino energy and the neutrino pseudorapidity, the more sizable are the differences between predictions from different PDF fits, emphasizing how far-forward neutrino experiments will be sensitive to large-$x$ PDFs. In this respect they will provide complementary constraints with respect to the electron-ion collider, which will explore the large-$x$ region in $ep$ and $eA$ collisions. On the other hand, we also observe that neutrinos with $\eta_\nu > 7.2$ and $E_\nu$ up to 700~GeV are mostly produced by $D$-mesons with $4.5 < y < 7.2$, as shown in Fig.~\ref{fig:promptnu}. This is the most relevant region contributing to the atmospheric flux of prompt neutrinos at $E_{\nu,\, lab} \sim \mathcal{O}$(PeV), signalling the connection between FPF and neutrino telescope measurements~\cite{Bai:2022xad}. 

\vspace{-0.2cm}
\section{Conclusions}
On the one hand, BSM phenomena - including neutrino oscillations - can be considered as possible backgrounds for the SM/QCD program at far-forward experiments. On the other hand, the effects produced by in\-te\-ractions induced by neutrinos or by BSM particles in the detectors can be disentangled by imposing specific kinematic cuts. The capability of distinguishing neutrinos and antineutrinos of dif\-fe\-rent flavours will help both the SM and the BSM programs. Using detectors covering different pseudorapidity ranges, with partial overlap among each other, will help to calibrate the detectors themselves, to check the robustness of their results, and to disentangle the role of perturbative and non-perturbative QCD effects. 

\subsection*{Acknowledgments}
I am grateful to my coauthors of 
ref.~\cite{Bai:2020ukz, Bai:2021ira, Bai:2022jcs, Bai:2022xad} and to my coauthors of ref.~\cite{Anchordoqui:2021ghd}, some of which attended and/or organized the ``Diffraction and Low-$x$ 2022'' Workshop, 
for joint work on the physics case of far-forward LHC ex\-pe\-ri\-ments and to M.~Benzke and K.~Marks for comments on 
this manuscript. 
This work is supported in part by the BMBF contract 05H21GUCCA.

\bibliographystyle{JHEPcustomized}
\bibliography{farforward}

\providecommand{\href}[2]{#2}\begingroup\raggedright\begin{thebibliography}{10}

\bibitem{DeRujula:1992sn}
A.~De~Rujula, E.~Fernandez and
  J.J.~Gomez-Cadenas\href{https://doi.org/10.1016/0550-3213(93)90427-Q}{\emph{Nucl.
  Phys. B} {\bfseries 405} (1993) 80}.

\bibitem{Park:2011gh}
H.~Park\href{https://doi.org/10.1007/JHEP10(2011)092}{\emph{JHEP} {\bfseries
  10} (2011) 092} [\href{https://arxiv.org/abs/1110.1971}{{\ttfamily
  1110.1971}}].

\bibitem{SNDLHC:2022ihg}
{\scshape SND@LHC} collaboration
  \href{https://arxiv.org/abs/2210.02784}{{\ttfamily 2210.02784}}.

\bibitem{FASER:2019dxq}
{\scshape FASER}
  collaboration\href{https://doi.org/10.1140/epjc/s10052-020-7631-5}{\emph{Eur.
  Phys. J. C} {\bfseries 80} (2020) 61}
  [\href{https://arxiv.org/abs/1908.02310}{{\ttfamily 1908.02310}}].

\bibitem{FASER:2022hcn}
{\scshape FASER} collaboration
  \href{https://arxiv.org/abs/2207.11427}{{\ttfamily 2207.11427}}.

\bibitem{FASER:2018eoc}
{\scshape FASER}
  collaboration\href{https://doi.org/10.1103/PhysRevD.99.095011}{\emph{Phys.
  Rev. D} {\bfseries 99} (2019) 095011}
  [\href{https://arxiv.org/abs/1811.12522}{{\ttfamily 1811.12522}}].

\bibitem{MammenAbraham:2020hex}
R.~Mammen~Abraham
  et~al.\href{https://doi.org/10.5281/zenodo.4059893}{\emph{Snowmass 2021 LoI}
  (2020) }.

\bibitem{Anchordoqui:2021ghd}
L.A.~Anchordoqui
  et~al.\href{https://doi.org/10.1016/j.physrep.2022.04.004}{\emph{Phys. Rept.}
  {\bfseries 968} (2022) 1} [\href{https://arxiv.org/abs/2109.10905}{{\ttfamily
  2109.10905}}].

\bibitem{Feng:2022inv}
J.L.~Feng et~al. \href{https://arxiv.org/abs/2203.05090}{{\ttfamily
  2203.05090}}.

\bibitem{Giunti:2022aea}
C.~Giunti, J.~Gruszko, B.~Jones, L.~Kaufman, D.~Parno and A.~Pocar
  \href{https://arxiv.org/abs/2209.03340}{{\ttfamily 2209.03340}}.

\bibitem{Bai:2020ukz}
W.~Bai, M.~Diwan, M.V.~Garzelli, Y.S.~Jeong and
  M.H.~Reno\href{https://doi.org/10.1007/JHEP06(2020)032}{\emph{JHEP}
  {\bfseries 06} (2020) 032}
  [\href{https://arxiv.org/abs/2002.03012}{{\ttfamily 2002.03012}}].

\bibitem{Foroughi-Abari:2020qar}
S.~Foroughi-Abari, F.~Kling and
  Y.-D.~Tsai\href{https://doi.org/10.1103/PhysRevD.104.035014}{\emph{Phys. Rev.
  D} {\bfseries 104} (2021) 035014}
  [\href{https://arxiv.org/abs/2010.07941}{{\ttfamily 2010.07941}}].

\bibitem{Batell:2021blf}
B.~Batell, J.L.~Feng and
  S.~Trojanowski\href{https://doi.org/10.1103/PhysRevD.103.075023}{\emph{Phys.
  Rev. D} {\bfseries 103} (2021) 075023}
  [\href{https://arxiv.org/abs/2101.10338}{{\ttfamily 2101.10338}}].

\bibitem{Basdew-Sharma:2022vya}
A.~Basdew-Sharma, A.~Pelloni, F.~Herzog and A.~Vogt
  \href{https://arxiv.org/abs/2211.16485}{{\ttfamily 2211.16485}}.

\bibitem{Zenaiev:2019ktw}
{\scshape PROSA}
  collaboration\href{https://doi.org/10.1007/JHEP04(2020)118}{\emph{JHEP}
  {\bfseries 04} (2020) 118}
  [\href{https://arxiv.org/abs/1911.13164}{{\ttfamily 1911.13164}}].

\bibitem{Alekhin:2018pai}
S.~Alekhin, J.~Bl\"umlein and
  S.~Moch\href{https://doi.org/10.1140/epjc/s10052-018-5947-1}{\emph{Eur. Phys.
  J. C} {\bfseries 78} (2018) 477}
  [\href{https://arxiv.org/abs/1803.07537}{{\ttfamily 1803.07537}}].

\bibitem{Dulat:2015mca}
S.~Dulat, T.-J.~Hou, J.~Gao, M.~Guzzi, J.~Huston, P.~Nadolsky
  et~al.\href{https://doi.org/10.1103/PhysRevD.93.033006}{\emph{Phys. Rev. D}
  {\bfseries 93} (2016) 033006}
  [\href{https://arxiv.org/abs/1506.07443}{{\ttfamily 1506.07443}}].

\bibitem{NNPDF:2017mvq}
{\scshape NNPDF}
  collaboration\href{https://doi.org/10.1140/epjc/s10052-017-5199-5}{\emph{Eur.
  Phys. J. C} {\bfseries 77} (2017) 663}
  [\href{https://arxiv.org/abs/1706.00428}{{\ttfamily 1706.00428}}].

\bibitem{Bai:2021ira}
W.~Bai, M.~Diwan, M.V.~Garzelli, Y.S.~Jeong, F.K.~Kumar and
  M.H.~Reno\href{https://doi.org/10.1007/JHEP06(2022)148}{\emph{JHEP}
  {\bfseries 06} (2022) 148}
  [\href{https://arxiv.org/abs/2112.11605}{{\ttfamily 2112.11605}}].

\bibitem{Bai:2022jcs}
W.~Bai, M.V.~Diwan, M.V.~Garzelli, Y.S.~Jeong, K.~Kumar and
  M.H.~Reno\href{https://doi.org/10.1016/j.jheap.2022.05.003}{\emph{JHEAp}
  {\bfseries 34} (2022) 212}
  [\href{https://arxiv.org/abs/2203.07212}{{\ttfamily 2203.07212}}].

\bibitem{Bai:2022xad}
W.~Bai, M.~Diwan, M.V.~Garzelli, Y.S.~Jeong, K.~Kumar and M.H.~Reno
  \href{https://arxiv.org/abs/2212.07865}{{\ttfamily 2212.07865}}.

\bibitem{Begel:2022kwp}
M.~Begel et~al. \href{https://arxiv.org/abs/2209.14872}{{\ttfamily
  2209.14872}}.

\end{thebibliography}\endgroup

\end{document}